\documentclass{article}
\usepackage{spconf,amsmath,graphicx}
\usepackage{multirow}
\usepackage{booktabs}
\usepackage{cite}
\usepackage[urlcolor=blue]{hyperref}
\usepackage{url}
\usepackage{bm}
\usepackage{graphicx}
\usepackage{subfigure}

\title{Neural Concatenative Singing Voice Conversion: Rethinking Concatenation-based Approach for One-shot Singing Voice Conversion}
\name{Binzhu Sha$^{1,*}$\thanks{$^{*}$ Equal contribution. This work was done when Binzhu Sha was an intern at ARC Lab, Tencent PCG.}, Xu Li$^{2,*, \dagger}$, Zhiyong Wu$^{1,3,\dagger}$\thanks{$^{\dagger}$ Corresponding authors.}, Ying Shan$^2$, Helen Meng$^{3}$}
\address{
    $^1$ Shenzhen International Graduate School, Tsinghua University, Shenzhen, China\\
    $^2$ ARC Lab, Tencent PCG\\
    $^3$ The Chinese University of Hong Kong, Hong Kong SAR, China\\
    \small{
        sbz22@mails.tsinghua.edu.cn, \{nelsonxli, yingsshan\}@tencent.com, zywu@sz.tsinghua.edu.cn, hmmeng@se.cuhk.edu.hk
    }
}

\ninept
\begin{document}
%
\maketitle
\begin{abstract}

Any-to-any singing voice conversion (SVC) is confronted with the challenge of ``timbre leakage'' issue caused by inadequate disentanglement between the content and the speaker timbre. To address this issue, this study introduces NeuCoSVC, a novel neural concatenative SVC framework. It consists of a self-supervised learning (SSL) representation extractor, a neural harmonic signal generator, and a waveform synthesizer. 
The SSL extractor condenses audio into fixed-dimensional SSL features, while the harmonic signal generator leverages linear time-varying filters to produce both raw and filtered harmonic signals for pitch information. The synthesizer reconstructs waveforms using SSL features, harmonic signals, and loudness information. 
During inference, voice conversion is performed by substituting source SSL features with their nearest counterparts from a matching pool which comprises SSL features extracted from the reference audio, while preserving raw harmonic signals and loudness from the source audio. By directly utilizing SSL features from the reference audio, the proposed framework effectively resolves the ``timbre leakage" issue caused by previous disentanglement-based approaches. 
Experimental results demonstrate that the proposed NeuCoSVC system outperforms the disentanglement-based speaker embedding approach in one-shot SVC across intra-language, cross-language, and cross-domain evaluations.

\end{abstract}

\section{Introduction}
\label{sec:intro}

The goal of singing voice conversion (SVC) is to transform the vocal characteristics of one singer into those of a target singer while preserving the content and melody of the song. With the advancement of neural network architectures, significant progress has been achieved in the task of SVC\cite{nachmani19_interspeech, Polyak2020, fastsvc, li2022hierarchical, hifisvc, diffsvc}, which brings extensive applications in human-computer interaction, entertainment, etc.

Current research in SVC predominantly adopts the encoder-decoder framework, where the encoder disentangles the singer's timbre from the linguistic content, and the decoder reconstruct the singing audio leveraging the distinct timbre information. Domain confusion loss\cite{nachmani19_interspeech} is utilized to ensure that the latent space of the encoder is singer-agnostic. Alternative approaches, such as obtaining singer-invariant content features from a linguistic extractor like an automatic speech recognition acoustic model\cite{Polyak2020, fastsvc, li2022hierarchical, zhou2023enhancing}, have also been explored. 
Diverse decoder architectures have been employed in SVC for singing audio synthesis, including auto-regressive models\cite{nachmani19_interspeech}, GANs\cite{fastsvc, li2022hierarchical, hifisvc}, and denoising diffusion probabilistic models\cite{diffsvc}.

Existing SVC models encounter challenges in any-to-any conversion, a task that involves adapting songs to the voices of unseen target speakers. The challenges arise from the need to generalize various vocal traits. Current models often use a speaker embedding vector, either from a look-up table (LUT)\cite{fastsvc, takahashi2021hierarchical} or a speaker verification model\cite{li2022hierarchical,zhang2020dursc,guo2022improving}, to capture speaker-specific traits.
However, the LUT method is limited by its predefined speaker set and struggles with unseen speakers. While a speaker verification model can overcome this, assuming the embedding vector contains all necessary vocal information is questionable.\cite{li2022hierarchical}.
Furthermore, effectively disentangling speaker characteristics from content remains challenging, potentially resulting in ``timbre leakage'', where source speakers' timbres partially remain in the converted audio\cite{wang21g_interspeech}.

\begin{figure*}[t]
\centering
	\includegraphics[width=0.9\textwidth]{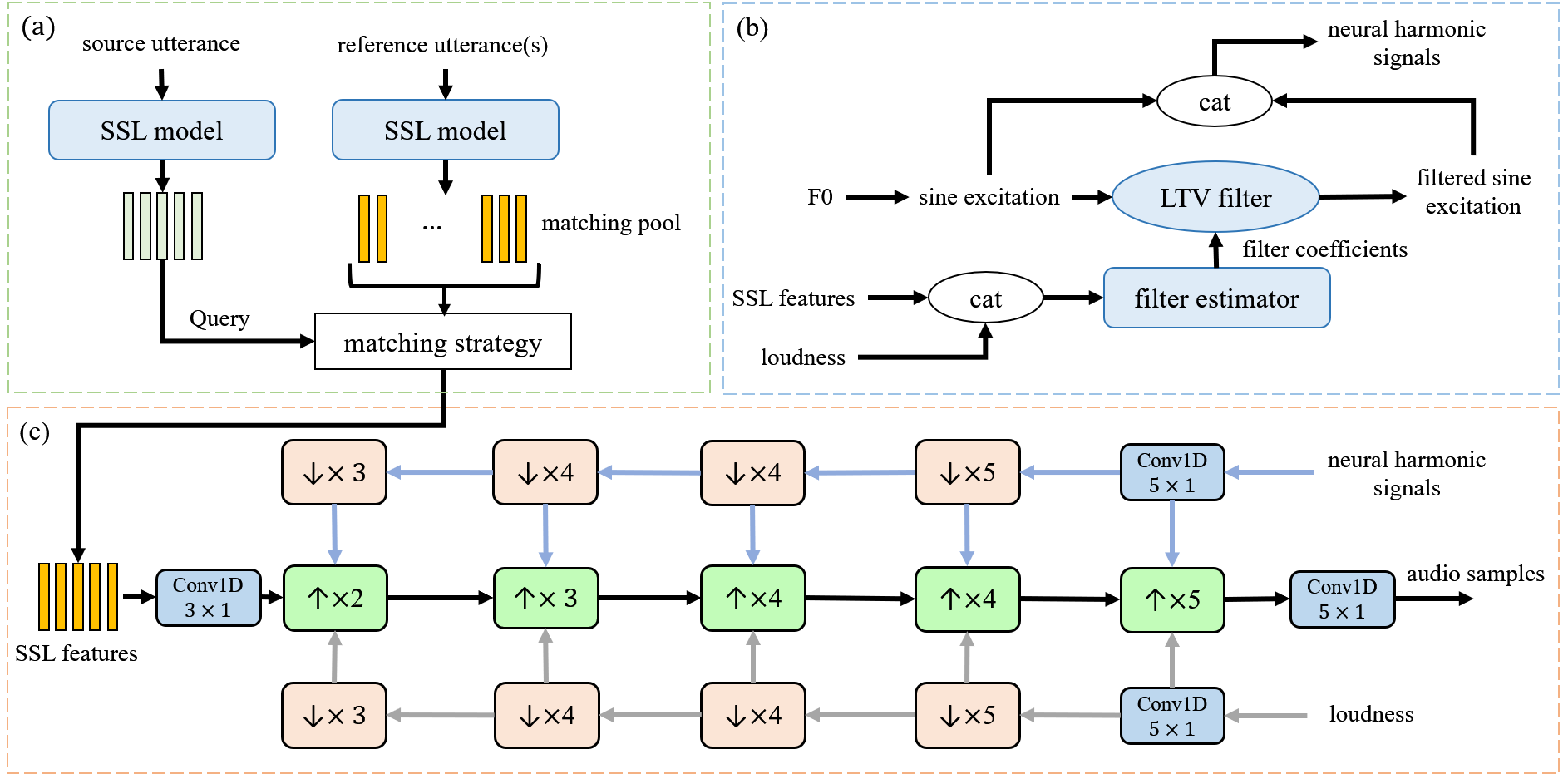}
	\caption{The structure of the proposed SVC system: (a) the SSL feature extracting and matching module; (b) the neural harmonic signal generator; (c) the audio synthesizer.}
	\label{fig: sys-architecture}
\end{figure*}

This study revisits the concatenation-based SVC to address timbre-content decoupling and timbre leakage challenges. Previous concatenation-based methods\cite{kei_fujii_2007_1054889, Cute_icassp14} using speech units from the target speaker suffer from quality distortion due to abrupt transitions between units. 
To overcome this, self-supervised learning (SSL) representations\cite{chen2022wavlm} containing phonetic and speaker information are used in the kNN-VC method\cite{knnvc_interspeech23}, which replaces speech units with SSL representations and directly synthesizes audio using a vocoder. During inference, SSL representations are extracted from the target audio to form a matching pool. Each frame of the source representation is then replaced with its nearest neighbors from the matching pool to create pre-matched representations for audio synthesis. This approach, which exclusively utilizes features derived from the target speaker, could potentially eliminate timbre leakage.

However, the absence of explicit pitch modeling in kNN-VC results in underperformance in SVC, leading to poor audio quality and out-of-tune outputs. Inspired by kNN-VC, this study introduces NeuCoSVC, a novel neural concatenation-based SVC approach. NeuCoSVC adopts the FastSVC architecture, known for its success in achieving high audio quality and voice similarity\cite{fastsvc, li2022hierarchical}. It incorporates explicit pitch modeling, a crucial feature for SVC, and effectively integrates pitch and loudness features through feature-wise linear modulation (FiLM)\cite{perez2018film}. Additionally, NeuCoSVC incorporates a linear time-varying (LTV) filter\cite{liu2020neural} to automatically adjust each harmonic component, further enhancing audio quality.

This work makes the following contributions: 
1) We develop a stronger baseline, NeuCo-HiFi-GAN, based on kNN-VC, by improving the matching strategy and integrating the neural source-filter module into HiFiGAN for pitch modeling.
2) We further propose a novel SVC framework, NeuCoSVC, which extends NeuCo-HiFi-GAN by leveraging the FastSVC architecture and the LTV module to enhance audio quality and voice similarity.
3) Extensive experiments are conducted across intra-language, cross-language, and cross-domain evaluations, demonstrating the promising performance of NeuCoSVC in all scenarios.
4) A duration study is conducted to assess the impact of reference audio duration on conversion performance, indicating that extending the duration beyond 60s yields diminishing returns.
Audio samples and codes are publicly available\footnote{https://github.com/thuhcsi/NeuCoSVC}.

\section{Proposed Method}
\label{sec:Method}

Our proposed NeuCoSVC system comprises three components: 1) the self-supervised learning (SSL) feature extraction and matching module, 2) the neural harmonic signal generator, and 3) the audio synthesizer. As depicted in Fig.~\ref{fig: sys-architecture}(a), the source audio is first processed by a pre-trained SSL model to obtain fixed-dimensional features encompassing linguistic and timbre information.
These SSL features are then matched with those from reference utterances to select semantically related features. The selected SSL features retain the source audio's semantic information while adopting the target person's timbre.
Fig.~\ref{fig: sys-architecture}(c) demonstrates the block-by-block up-sampling of these pre-matched SSL features into audio samples via the audio synthesizer.
Additionally, we employ linear time-varying (LTV) filters to derive filtered harmonic signals for pitch reconstruction, as shown in Fig.\ref{fig: sys-architecture}(b), which has been proven to effectively enhance the quality and similarity of the converted audio\cite{guo2022improving}. 
The following sections will explore these sub-modules in detail.

\subsection{The SSL feature extraction and matching module}
\label{subsec:ssl-feature-matching}
The SSL feature matching module consists of two stages: 1) extracting compact features from audio, and 2) replacing the extracted features in the source utterance with those from the reference utterances. In the first stage, we follow the methodology of kNN-VC and adopt the pre-trained WavLM-Large encoder \cite{chen2022wavlm} to extract SSL features from audio. WavLM is a self-supervised model that jointly learns masked speech prediction and denoising, enabling it to tackle full-stack downstream speech tasks such as speaker verification, speech separation, and speech recognition. 

It has been discovered that the 6th layer of WavLM efficiently maps frames of the same phoneme closer in feature space compared to frames of different phonemes. Notably, this layer also preserves the speaker's timbre information, making it valuable for voice conversion\cite{knnvc_interspeech23}.
Hence, replacement on these features ensures that the output voice retains the source content while only altering speaker characteristics. 
Regarding the matching strategy, we employ the kNN method, searching for the $K$ nearest SSL features in the reference matching pool and averaging these $K$ features for replacement. 
To improve the accuracy of the matching process, we employ the mean of the last 5 layers from WavLM-Large for matching, while utilizing the 6th layer for synthesis. This decision is motivated by the fact that the last 5 layers contain more discriminative content information\cite{chen2022wavlm, lin2022utility}, thereby enhancing the matching accuracy.

\subsection{The neural harmonic signal generator}
\label{subsec: neural-harmonic-signals-generator}
To improve the quality and similarity of the converted audio, we utilize neural harmonic signals to represent the pitch information. This involves up-sampling the frame-level $f_{0}$ features into audio-level and deriving the sine excitation $p[n]$ by Eq.\ref{eq: sine_excitation} and Eq.\ref{eq: sine_excitation_K}.
\begin{align}
    p[n] &= \begin{cases}
     \sum_{k=1}^{K} \cos (\frac{2\pi k\sum_{i=0}^{n} f_{0}[i]}{f_{s}}), & f_{0}[n] > 0 \\
     0, & f_{0}[n] = 0
    \end{cases} \label{eq: sine_excitation} \\
    K &= \lfloor \frac{f_{s}}{2f_{0}[n]} \rfloor \label{eq: sine_excitation_K}
\end{align}
where $k \in \{1, 2, ..., K\}$ and $n$ are the indexes of harmonics and time, respectively. $K$ is the largest number of harmonics that can be achieved corresponding to the $f_{0}[n]$. $f_{s}$ is the audio sampling rate.

Next, LTV filters \cite{liu2020neural} are applied to adjust the amplitudes of different harmonic components based on condition features at each time, as shown in Eq.\ref{eq: neural_harmonic_filtered}.
\begin{align}
    \widetilde p[n] = h_{1}[n] \ast p[n] + h_{2}[n] \ast z[n] \label{eq: neural_harmonic_filtered}
\end{align}
where $h_{1}[n]$ and $h_{2}[n]$ are the LTV filters operating on the $p[n]$ and $z[n]$, respectively. The ``$\ast$'' is the convolution operation. The $z[n] \sim \mathcal{N}(0, 0.03^2)$ is a signal sequence randomly generated from a Gaussian distribution. The coefficients of $h_{1}[n]$ and $h_{2}[n]$ are estimated by a neural filter estimator, conditioned on the concatenation of the SSL features and the loudness features.

Finally, we concatenate the raw sine excitation and the filtered excitation signals to form the neural harmonic signals, which are fed into the audio synthesizer for pitch conditioning. Eq.\ref{eq: neural_harmonic_both} illustrates this operation, where $\oplus$ represents the concatenation operation.
\begin{align}
    s[n] = p[n] \oplus \widetilde p[n] \label{eq: neural_harmonic_both}
\end{align}

\subsection{The audio synthesizer}
\label{subsec:audio-synthesizer}
The audio synthesizer comprises one up-sampling stream and two down-sampling streams. The up-sampling stream, with five up-sampling blocks, gradually transforms the SSL features into audio samples. The two down-sampling streams down-sample the pitch and the loudness information into the corresponding scale at each block, respectively. Then, the pitch and loudness are integrated into the up-sampling block via a ``FiLM'' module, as described in \cite{li2022hierarchical}.

\section{Experimental Setup}
\label{sec:expt-setup}

\subsection{Dataset}
Experiments are conducted on the OpenSinger dataset \cite{opensinger}, consisting of 50 hours of high-quality Chinese singing voices recorded in professional studios. It includes 28 male and 48 female singers, with audio saved in wav format at the sampling rate of 44.1 kHz. Singing utterances of two male and two female singers are retained as the test set, and the remaining utterances are randomly divided into train-validation sets at a ratio of 9:1. This work also assesses the cross-lingual SVC performance using the NUS-48E dataset \cite{nus48E}, which features six male and six female English singers, each producing 4 English songs, with a total duration of about 10 minutes per singer. Two male and two female singers from the NUS-48E dataset are randomly selected for the cross-lingual SVC evaluation.

\subsection{Preprocessing \& Model configurations}

Pitch values are extracted by taking the median of three methods (PYIN\footnote{\href{https://github.com/librosa/librosa}{https://github.com/librosa/librosa}}, REAPER\footnote{\href{https://github.com/google/REAPER}{https://github.com/google/REAPER}} and Parselmouth\footnote{\href{https://github.com/YannickJadoul/Parselmouth}{https://github.com/YannickJadoul/Parselmouth}}). The A-weighting mechanism of the power spectrum \cite{measuring_noise} is adopted as the loudness feature. Pitch and loudness features are extracted with a hop size of 10ms. Notably, to fit within the target speaker's vocal range, the source pitch values are multiplied by a shift factor during conversion. The shift factor is the ratio of the median pitch value in the target audio to that in the source audio.

The pre-trained WavLM-Large\cite{chen2022wavlm} is utilized to extract self-supervised learning (SSL) features, generating a single vector for every 20 ms of 16 kHz audio. 
To align SSL features with the pitch and loudness, SSL features of each frame are duplicated twice.
The k-nearest neighbor method with $k=4$ is applied for the matching strategy, taking the cosine similarity as the distance metric.
The setting of the audio synthesizer strictly follows \cite{li2022hierarchical} except for the up/down-sampling factors, as shown in Fig.~\ref{fig: sys-architecture}(c).

\subsection{Training configurations}
The proposed system is trained on the OpenSinger, except that the SSL model parameters are fixed during training.
The audio synthesizer accepts 1024-dimensional SSL features, neural harmonic signals and loudness as inputs, generating 24 kHz singing audios.
Following \cite{knnvc_interspeech23}, we prepare the training set by generating the pre-matched SSL features, which are produced by replacing each original feature with the average of its k nearest neighbors within the same speaker's feature set. This pre-matching strategy bridges the gap between training and inference, improving the intelligibility of the converted audio.
The ADAM algorithm with an initial learning rate of 0.001 is used as the optimizer. The least-squares GAN \cite{GAN} loss and multi-scale STFT loss \cite{Parallel_Wavegan} are adopted during training.

\begin{table*}
\centering
\caption{Subjective and objective evaluation results. MOS results are reported with 95\% confidence intervals. Voice similarity is measured using cosine similarity of pretrained ASV model-derived speaker embedding vectors, with Ground Truth denoting source-target. The number of parameters of the corresponding audio synthesizer is shown in the brackets after each model.}
\label{tab:experiment_results}
\begin{tabular}{cccccccc}
\hline
\multirow{3}{*}{Model} & \multicolumn{6}{c}{MOS results}                                                                                             & \multirow{2}{*}{Voice similarity} \\ \cline{2-7}
                       & \multicolumn{3}{c}{Naturalness}                              & \multicolumn{3}{c}{Similarity}                               &                                   \\ \cline{2-8}
                       & IL                 & CL                 & CD                 & IL                 & CL                 & CD                 & Source/Converted-target                  \\ \hline
Ground Truth           & 4.47±0.10          & 4.58±0.08          & 4.78±0.06          & \textbackslash{}   & \textbackslash{}   & \textbackslash{}   & 0.16                              \\
SpkEmb-FastSVC         & 3.42±0.13          & 3.56±0.12          & 3.33±0.12          & 3.38±0.12          & 3.45±0.12          & 3.27±0.12          & 0.35                              \\
NeuCo-HiFi-GAN (16.54M)        & 3.62±0.12          & 3.46±0.13          & 3.10±0.13          & 3.35±0.12          & 3.22±0.12          & 2.80±0.13          & 0.44                              \\
NeuCo-FastSVC (12.38M)         & 3.65±0.13          & 3.73±0.12          & 3.43±0.13          & 3.52±0.11          & \textbf{3.61±0.12} & 3.41±0.12          & 0.52                              \\
NeuCoSVC (14.45M)              & \textbf{3.95±0.11} & \textbf{3.90±0.12} & \textbf{3.56±0.12} & \textbf{3.59±0.12} & 3.59±0.12          & \textbf{3.48±0.11} & \textbf{0.57}                     \\ \hline
\end{tabular}
\end{table*}

\subsection{Baselines}

We compare the proposed NeuCoSVC system with three baseline systems: SpkEmb-FastSVC, NeuCo-HiFi-GAN, and NeuCo-FastSVC. 
SpkEmb-FastSVC utilizes speaker embedding vectors extracted from ECAPA-TDNN \cite{ecapa-tdnn} for timbre information, while linguistic features are extracted using a pre-trained automatic speech recognition (ASR) conformer-based encoder\cite{zhang2022wenetspeech}.
In contrast, NeuCo-FastSVC leverages the neural concatenative method introduced in this work. Both systems share the same audio synthesizer. 
NeuCo-HiFiGAN is developed based on kNN-VC, employing the Hifi-GAN\cite{hifigan} vocoder with a neural source-filter module \cite{nsf} to explicitly model pitch during audio synthesis. 

\section{Experiment Results}
\label{sec:expt-rst}

We conduct any-to-any singing voice conversion experiments in three different scenarios: intra-language (IL) and cross-language (CL) conversions for in-domain tests, and intra-language conversions for the cross-domain evaluation. In in-domain SVC, target speakers' singing voices are employed, whereas, in cross-domain SVC, their speech voices serve as input.
We randomly select twelve audio segments from the OpenSinger dataset as source audio. Each segment is converted into the singing voice of four OpenSinger test singers and four from NUS48E for the in-domain test, with approximately ten minutes of reference audio per conversion. Additionally, four Chinese speech segments, delivered by different speakers and lasting around half a minute, are used for cross-domain evaluation.
The standard 5-scale mean opinion score (MOS) test is conducted to evaluate the naturalness and the voice similarity. Eighteen Chinese speakers proficient in English participated in subjective evaluations.

\subsection{Subjective evaluation}

Table.\ref{tab:experiment_results} presents the results of subjective evaluations. NeuCo-FastSVC demonstrates better performance over SpkEmb-FastSVC in terms of both naturalness and similarity, showcasing the advantages of the concatenative approach over the disentanglement-based speaker embedding approach in one-shot SVC. We also conduct ablation studies to validate the efficacy of integrating the FastSVC architecture and neural harmonic filters. When comparing NeuCo-FastSVC with NeuCo-HiFi-GAN, the former outperforms in both aspects, indicating that the FastSVC architecture is better suited for SVC by effectively reconstructing singing voice from SSL features. Additionally, in the comparison between NeuCoSVC and NeuCo-FastSVC, while maintaining a similar level of timbre similarity, NeuCoSVC achieves superior naturalness, highlighting the effectiveness of incorporating neural harmonic filters.
In the cross-language scenario, NeuCoSVC achieves performance comparable to intra-language conversions, which could be attributed to the similarity of fundamental phonemes between different languages, indicating its applicability for conversions between diverse languages.
In the cross-domain scenario, all models suffer from performance degradation, likely due to the short reference audio duration, differences in pitch range and the gap between speech and singing audio. Nevertheless, NeuCoSVC demonstrates robustness and broader applicability, yielding favorable results.

\subsection{Objective evaluation}

We use an automatic speaker verification (ASV) model to assess the accuracy of singing conversion results as an objective metric. Specifically, we train an ECAPA-TDNN \cite{desplanques2020ecapa} model on a large-scale combined dataset of VoxCeleb1\cite{nagrani2017voxceleb}, VoxCeleb2\cite{chung2018voxceleb2}, CN-Celeb1\cite{fan2020cn}, and CN-Celeb2 \cite{li2022cn}, totaling 3200+ hours of speech from 10k+ speakers. 
We employ the pre-trained ECAPA-TDNN model to compute speaker embedding vectors for each source audio, target audio, and converted audio. We then calculate the cosine similarity between the speaker embedding vectors of source and target audios, as well as between the converted audio and the target audio. Experimental results, as shown in Table.\ref{tab:experiment_results}, indicate that the neural concatenative-based singing voice conversion model achieves higher similarity compared to the model using speaker embedding vectors.

\subsection{Duration Study}

\begin{figure}[!t]
\centering
	\includegraphics[width=8.6cm]{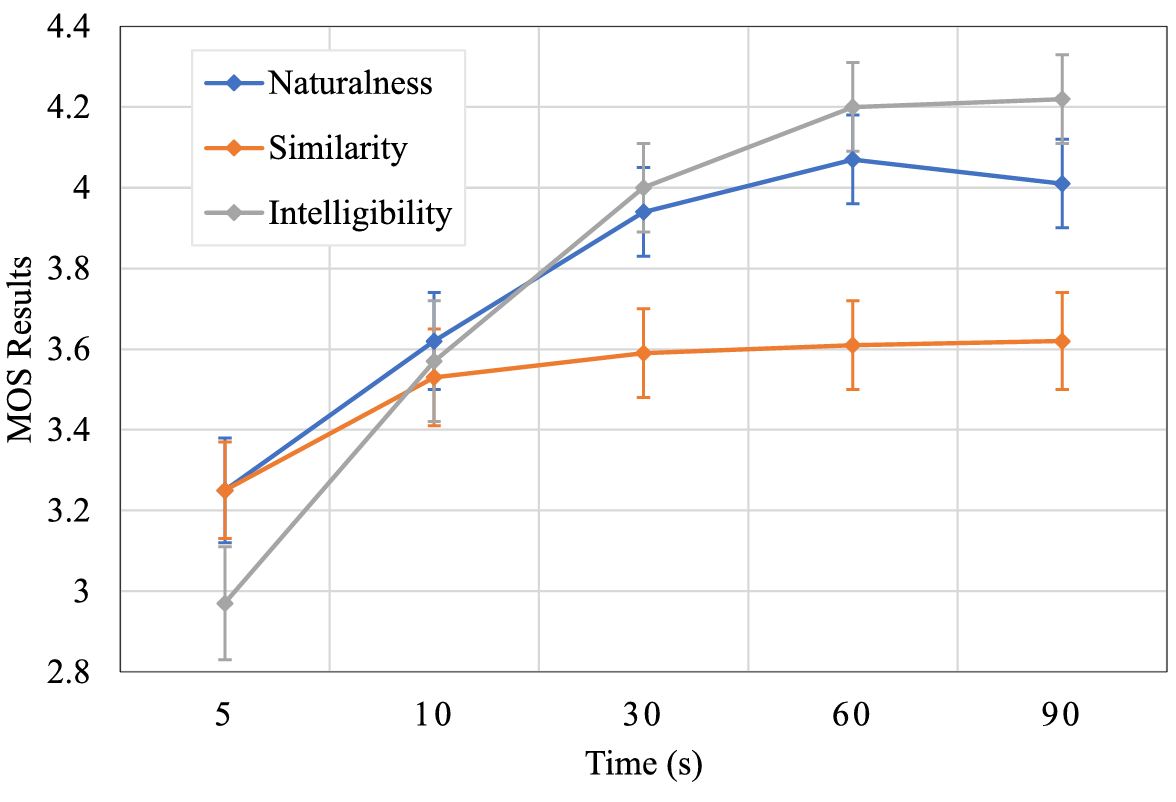}
	\caption{Experimental results of the duration study. MOS results are calculated with 95\% confidence intervals.}
	\label{fig: duration_study}
\end{figure}

\begin{table}
\caption{the number of unique phones/biphones in reference audio.}
\label{tab:duration_study}
\centering
\begin{tabular}{lccccc}\hline
          &  5s   &  10s  &  30s  &  60s  & 90s   \\\hline
phonemes  &  14   &  23	&  38	&  46	& 49 \\\hline                
\end{tabular}
\end{table}

Since the SSL features are exclusively derived from the matching pool extracted from reference audio, the phonemic richness within the reference audio significantly affects the quality of the singing audio. 
To assess the impact of reference audio duration on conversion quality, we conduct tests using singing audio of varying lengths from the OpenSinger dataset.
Evaluators rated the converted audio based on intelligibility, singing naturalness, and timbre similarity.
As depicted in Fig.\ref{fig: duration_study}, in extreme cases, e.g. 5 seconds of reference audio, the synthesizer struggles to generate higher-quality singing audio due to the limited phonemes coverage. However, as the duration of reference audio reaches 1 minute, all singing audio exhibits good intelligibility and naturalness. 
Further extending reference audio duration yields diminishing improvements in audio quality, as much of the phonemic content is already captured. 
Statistical analysis on the number of unique phones/biphones in reference audios of varying durations supports this observation, as presented in Table.\ref{tab:duration_study}.

\section{Conclusion}
\label{sec:conclusion}
In this work, we introduce NeuCoSVC, a neural concatenative SVC architecture for any-to-any conversion. 
Our approach effectively alleviates the timbre leakage issue in previous disentanglement-based approaches.
Experimental results demonstrate that the proposed NeuCoSVC outperforms the baseline method in terms of both naturalness and similarity across intro-language, cross-language, and cross-domain scenarios.
The ablation studies are carried out to confirm the effectiveness of components in the NeuCoSVC model.

\section{Acknowledgment}
\label{sec:acknowledgment}
This work is supported by National Natural Science Foundation of China (62076144), National Social Science Foundation of China (13\&ZD189), Shenzhen Science and Technology Program (WDZC20220816140515001, JCYJ20220818101014030) and Tencent AI Lab Rhino-Bird Focused Research Program (RBFR2023015).




\bibliographystyle{IEEEtran}
\bibliography{strings,refs}

\begin{thebibliography}{10}
\providecommand{\url}[1]{#1}
\csname url@samestyle\endcsname
\providecommand{\newblock}{\relax}
\providecommand{\bibinfo}[2]{#2}
\providecommand{\BIBentrySTDinterwordspacing}{\spaceskip=0pt\relax}
\providecommand{\BIBentryALTinterwordstretchfactor}{4}
\providecommand{\BIBentryALTinterwordspacing}{\spaceskip=\fontdimen2\font plus
\BIBentryALTinterwordstretchfactor\fontdimen3\font minus
  \fontdimen4\font\relax}
\providecommand{\BIBforeignlanguage}[2]{{%
\expandafter\ifx\csname l@#1\endcsname\relax
\typeout{** WARNING: IEEEtran.bst: No hyphenation pattern has been}%
\typeout{** loaded for the language `#1'. Using the pattern for}%
\typeout{** the default language instead.}%
\else
\language=\csname l@#1\endcsname
\fi
#2}}
\providecommand{\BIBdecl}{\relax}
\BIBdecl

\bibitem{nachmani19_interspeech}
E.~Nachmani and L.~Wolf, ``{Unsupervised Singing Voice Conversion},'' in
  \emph{Proc. Interspeech}, 2019, pp. 2583--2587.

\bibitem{Polyak2020}
A.~Polyak, L.~Wolf, Y.~Adi, and Y.~Taigman, ``{Unsupervised Cross-Domain
  Singing Voice Conversion},'' in \emph{Proc. Interspeech}, 2020, pp. 801--805.

\bibitem{fastsvc}
S.~Liu, Y.~Cao, N.~Hu, D.~Su, and H.~Meng, ``Fastsvc: Fast cross-domain singing
  voice conversion with feature-wise linear modulation,'' in \emph{ICME}, 2021,
  pp. 1--6.

\bibitem{li2022hierarchical}
X.~Li, S.~Liu, and Y.~Shan, ``A hierarchical speaker representation framework
  for one-shot singing voice conversion,'' \emph{INTERSPEECH}, 2022.

\bibitem{hifisvc}
Y.~Zhou and X.~Lu, ``Hifi-svc: Fast high fidelity cross-domain singing voice
  conversion,'' in \emph{ICASSP}, 2022, pp. 6667--6671.

\bibitem{diffsvc}
S.~Liu, Y.~Cao, D.~Su, and H.~Meng, ``Diffsvc: A diffusion probabilistic model
  for singing voice conversion,'' in \emph{IEEE Automatic Speech Recognition
  and Understanding Workshop (ASRU)}, 2021, pp. 741--748.

\bibitem{zhou2023enhancing}
S.~Zhou, X.~Li, Z.~Wu, Y.~Shan, and H.~Meng, ``Enhancing the vocal range of
  single-speaker singing voice synthesis with melody-unsupervised
  pre-training,'' in \emph{ICASSP}.\hskip 1em plus 0.5em minus 0.4em\relax
  IEEE, 2023, pp. 1--5.

\bibitem{takahashi2021hierarchical}
N.~Takahashi, M.~K. Singh, and Y.~Mitsufuji, ``Hierarchical disentangled
  representation learning for singing voice conversion,'' in
  \emph{International Joint Conference on Neural Networks (IJCNN)}.\hskip 1em
  plus 0.5em minus 0.4em\relax IEEE, 2021, pp. 1--7.

\bibitem{zhang2020dursc}
L.~Zhang, C.~Yu, H.~Lu, C.~Weng, C.~Zhang, Y.~Wu, X.~Xie, L.~Zijin, and D.~Yu,
  ``Durian-sc: Duration informed attention network based singing voice
  conversion system,'' in \emph{Proc. Interspeech 2020}, 10 2020, pp.
  1231--1235.

\bibitem{guo2022improving}
H.~Guo, Z.~Zhou, F.~Meng, and K.~Liu, ``Improving adversarial waveform
  generation based singing voice conversion with harmonic signals,''
  \emph{ICASSP}, 2022.

\bibitem{wang21g_interspeech}
Z.~Wang, X.~Zhou, F.~Yang, T.~Li, H.~Du, L.~Xie, W.~Gan, H.~Chen, and H.~Li,
  ``{Enriching Source Style Transfer in Recognition-Synthesis Based
  Non-Parallel Voice Conversion},'' in \emph{Proc. Interspeech 2021}, 2021, pp.
  831--835.

\bibitem{kei_fujii_2007_1054889}
K.~Fujii, J.~Okawa, and K.~Suigetsu, ``{High-Individuality Voice Conversion
  Based on Concatenative Speech Synthesis},'' Nov. 2007.

\bibitem{Cute_icassp14}
Z.~Jin, A.~Finkelstein, S.~DiVerdi, J.~Lu, and G.~J. Mysore, ``Cute: A
  concatenative method for voice conversion using exemplar-based unit
  selection,'' in \emph{ICASSP}, 2016, pp. 5660--5664.

\bibitem{chen2022wavlm}
S.~Chen, C.~Wang, Z.~Chen, Y.~Wu, S.~Liu, Z.~Chen, J.~Li, N.~Kanda,
  T.~Yoshioka, X.~Xiao \emph{et~al.}, ``Wavlm: Large-scale self-supervised
  pre-training for full stack speech processing,'' \emph{IEEE Journal of
  Selected Topics in Signal Processing}, vol.~16, no.~6, pp. 1505--1518, 2022.

\bibitem{knnvc_interspeech23}
M.~Baas, B.~{van Niekerk}, and H.~Kamper, ``{Voice Conversion With Just Nearest
  Neighbors},'' in \emph{Proc. INTERSPEECH}, 2023, pp. 2053--2057.

\bibitem{perez2018film}
E.~Perez, F.~Strub, H.~De~Vries, V.~Dumoulin, and A.~Courville, ``Film: Visual
  reasoning with a general conditioning layer,'' in \emph{Proceedings of the
  AAAI conference on artificial intelligence}, vol.~32, 2018.

\bibitem{liu2020neural}
Z.~Liu, K.~Chen, and K.~Yu, ``Neural homomorphic vocoder.'' in
  \emph{INTERSPEECH}, 2020, pp. 240--244.

\bibitem{lin2022utility}
G.-T. Lin, C.-L. Feng, W.-P. Huang, Y.~Tseng, T.-H. Lin, C.-A. Li, H.-y. Lee,
  and N.~G. Ward, ``On the utility of self-supervised models for
  prosody-related tasks,'' in \emph{2022 IEEE Spoken Language Technology
  Workshop (SLT)}, 2023, pp. 1104--1111.

\bibitem{opensinger}
R.~Huang, F.~Chen, Y.~Ren, J.~Liu, C.~Cui, and Z.~Zhao, ``Multi-singer: Fast
  multi-singer singing voice vocoder with a large-scale corpus,'' in
  \emph{Proceedings of the 29th ACM International Conference on Multimedia},
  ser. MM '21, 2021, p. 3945–3954.

\bibitem{nus48E}
Z.~Duan, H.~Fang, B.~Li, K.~C. Sim, and Y.~Wang, ``The nus sung and spoken
  lyrics corpus: A quantitative comparison of singing and speech,'' in
  \emph{Asia-Pacific Signal and Information Processing Association Annual
  Summit and Conference}, 2013, pp. 1--9.

\bibitem{measuring_noise}
C.~Meyer-Bisch, ``Measuring noise,'' \emph{Medecine sciences : M/S}, vol.~21,
  no.~5, p. 546—550, May 2005.

\bibitem{GAN}
X.~Mao, Q.~Li, H.~Xie, R.~Y. Lau, Z.~Wang, and S.~P. Smolley, ``Least squares
  generative adversarial networks,'' in \emph{IEEE International Conference on
  Computer Vision (ICCV)}, 2017, pp. 2813--2821.

\bibitem{Parallel_Wavegan}
R.~Yamamoto, E.~Song, and J.-M. Kim, ``Parallel wavegan: A fast waveform
  generation model based on generative adversarial networks with
  multi-resolution spectrogram,'' in \emph{ICASSP}, 2020, pp. 6199--6203.

\bibitem{ecapa-tdnn}
B.~Desplanques, J.~Thienpondt, and K.~Demuynck, ``{ECAPA-TDNN: Emphasized
  Channel Attention, Propagation and Aggregation in TDNN Based Speaker
  Verification},'' in \emph{Proc. Interspeech}, 2020, pp. 3830--3834.

\bibitem{zhang2022wenetspeech}
B.~Zhang, H.~Lv, P.~Guo, Q.~Shao, C.~Yang, L.~Xie, X.~Xu, H.~Bu, X.~Chen,
  C.~Zeng \emph{et~al.}, ``Wenetspeech: A 10000+ hours multi-domain mandarin
  corpus for speech recognition,'' in \emph{ICASSP}.\hskip 1em plus 0.5em minus
  0.4em\relax IEEE, 2022, pp. 6182--6186.

\bibitem{hifigan}
J.~Kong, J.~Kim, and J.~Bae, ``Hifi-gan: Generative adversarial networks for
  efficient and high fidelity speech synthesis,'' in \emph{Advances in Neural
  Information Processing Systems}, vol.~33, 2020, pp. 17\,022--17\,033.

\bibitem{nsf}
X.~Wang, S.~Takaki, and J.~Yamagishi, ``Neural source-filter waveform models
  for statistical parametric speech synthesis,'' \emph{IEEE/ACM Transactions on
  Audio, Speech, and Language Processing}, vol.~28, pp. 402--415, 2020.

\bibitem{desplanques2020ecapa}
B.~Desplanques, J.~Thienpondt, and K.~Demuynck, ``Ecapa-tdnn: Emphasized
  channel attention, propagation and aggregation in tdnn based speaker
  verification,'' \emph{INTERSPEECH}, 2020.

\bibitem{nagrani2017voxceleb}
A.~Nagrani, J.~S. Chung, and A.~Zisserman, ``Voxceleb: a large-scale speaker
  identification dataset,'' \emph{arXiv preprint arXiv:1706.08612}, 2017.

\bibitem{chung2018voxceleb2}
J.~S. Chung, A.~Nagrani, and A.~Zisserman, ``Voxceleb2: Deep speaker
  recognition,'' \emph{arXiv preprint arXiv:1806.05622}, 2018.

\bibitem{fan2020cn}
Y.~Fan, J.~Kang, L.~Li, K.~Li, H.~Chen, S.~Cheng, P.~Zhang, Z.~Zhou, Y.~Cai,
  and D.~Wang, ``Cn-celeb: a challenging chinese speaker recognition dataset,''
  in \emph{ICASSP}.\hskip 1em plus 0.5em minus 0.4em\relax IEEE, 2020, pp.
  7604--7608.

\bibitem{li2022cn}
L.~Li, R.~Liu, J.~Kang, Y.~Fan, H.~Cui, Y.~Cai, R.~Vipperla, T.~F. Zheng, and
  D.~Wang, ``Cn-celeb: multi-genre speaker recognition,'' \emph{Speech
  Communication}, vol. 137, pp. 77--91, 2022.

\end{thebibliography}

\end{document}